\documentclass[%
 reprint,
 amsmath,amssymb,
 aps,
]{revtex4-2}

\usepackage{graphicx}
\usepackage{dcolumn}
\usepackage{bm}
\usepackage{multirow}

\begin{document}

\preprint{APS/123-QED}

\title{Lattice Thermal Conductivity of 8‐16‐4(Sun)-Graphyne from\\Reverse Nonequilibrium Molecular Dynamics Simulations}

\author{Isaac de Mac\^edo Felix}
\affiliation{Department of Physics, Federal University of Pernambuco, Recife, Pernambuco, Brazil.}
\author{Raphael Matozo Tromer}
\affiliation{Department of Applied Physics, State University of Campinas, Campinas, São Paulo, Brazil.}
\author{Leonardo Dantas Machado}
\affiliation{Department of Physics, Federal University of Rio Grande do Norte, Natal, Rio Grande do Norte, Brazil.}
\author{Douglas Soares Galvão}
\affiliation{Department of Applied Physics and Center for Computational Engineering and Sciences, State University of Campinas, Campinas, São Paulo, Brazil.}
\author{Luiz Antonio Ribeiro, Jr}
\affiliation{University of Brasília, Institute of Physics, Brasília, Federal District, Brazil.}
\author{Marcelo Lopes Pereira, Jr}
\affiliation{University of Bras\'{i}lia, College of Technology, Department of Electrical Engineering, Bras\'{i}lia, Federal District, Brazil.}

\date{\today}

\begin{abstract}
The thermal conductivity of two-dimensional (2D) materials is critical in determining their suitability for several applications, from electronics to thermal management. In this study, we have used Molecular Dynamics (MD) simulations to investigate the thermal conductivity and phononic properties of 8-16-4(Sun)-Graphyne, a recently proposed 2D carbon allotrope. The thermal conductivity was estimated using reverse non-equilibrium MD simulations following the Müller-Plathe approach, revealing a strong dependence on system size. Phonon dispersion calculations confirm the stability of Sun-GY while also showing a significant decrease in thermal conductivity compared to graphene. This decrease is attributed to acetylenic bonds, which enhance phonon scattering. Spectral analysis further revealed that Sun-GY exhibits lower phonon group velocities and increased phonon scattering, mainly due to interactions between acoustic and optical modes. Sun-GY presents an intrinsic thermal conductivity of approximately 24.6 W/mK, much lower than graphene, making it a promising candidate for applications that require materials with reduced thermal transport properties.
\end{abstract}

\maketitle


\section{Introduction} 

Thermal transport in nanomaterials has attracted growing interest from the scientific community and industry due to its critical role in various technological applications \cite{qiu2020review}. When a material’s dimensions are reduced to the nanometric scale, heat conduction --- primarily governed by phonons --- begins to exhibit unique behaviors compared to those observed in macroscopic materials \cite{cahill2002thermometry}. Factors such as interfaces, edges, and quantum confinement alter phonon propagation, resulting in significant variations in thermal conductivity depending on the material’s structure and topology \cite{xie2023brief}. These distinctive thermal transport properties make nanomaterials particularly promising for emerging technologies, including high-performance electronic devices, where efficient heat management is essential, and thermoelectric materials with improved energy efficiency \cite{yang2018high}. Detailed studies of these thermal properties are crucial for unlocking the full potential of nanomaterials in future innovations \cite{prasher2006thermal}.

Two-dimensional (2D) materials have attracted significant attention due to their potential applications in nanoelectronic devices and advanced thermal management systems \cite{song2018two}. Among these, carbon is a highly versatile element, pivotal in advancing nanotechnology \cite{dai2012carbon}. There is a renewed interest in carbon-based materials following the synthesis of graphene in 2004 \cite{novoselov2004electric} due to its remarkable thermal conductivity and other unique physicochemical properties \cite{idumah2016emerging}. This breakthrough led to the exploration of other carbon-based structures \cite{idumah2016emerging}, including the graphyne (GY) family, which exhibit a variety of atomic arrangements and diverse physical properties \cite{baughman1987structure,wang2016,desyatkin2022scalable}. GYs are topologically formed by introducing acetylene linkages between carbon atoms in graphene hexagonal lattices or other carbon allotropes. Both $sp^2$-hybridized allotropes and their $sp$-$sp^2$ analogs continue to be the focus of extensive research due to their potential in novel nanotechnological applications \cite{li2021effect,peng2014new,kang2018graphyne}.

As mentioned above, graphene was experimentally realized in 2004. Significant progress in carbon nanomaterials has been attained since the synthesis of fullerenes \cite{kroto1985c60} and carbon nanotubes \cite{iijima1991helical}. Theoretical predictions for the GY family even predate graphene, with the first work reported by Baughman and collaborators in 1987 \cite{baughman1987structure}. Over the years, considerable work has been invested in developing synthesis routes for these 2D materials \cite{alam2021synthesis}. The first GY monolayer, graphdiyne, was successfully synthesized in 2010 \cite{li2010architecture}, followed by $\gamma$-graphyne, which was synthesized via multiple routes between 2018 and 2022 \cite{li2018synthesis,hu2022synthesis,desyatkin2022scalable,he2023one}. Another notable example is the holey graphyne (HGY), which has also been synthesized \cite{liu2022constructing}. These advances extend beyond GYs, with other theoretically predicted materials being synthesized recently, such as PHA-Graphene \cite{fan2019nanoribbons,wang2015phagraphene}, Biphenylene \cite{fan2021biphenylene,hudspeth2010electronic}, and 2D fullerene networks \cite{hou2022synthesis,berber2004rigid}.

As mentioned, the GY family is broad, encompassing various phases \cite{kang2018graphyne}. Among the most well-known are the $\alpha$, $\beta$, $\gamma$, $\delta$, $\lambda$, and 6,6,12 phases \cite{narang2023review}. The thermal conduction properties of these GY nanomaterials have been extensively studied. For example, Wang \textit{et al.} used the Reactive Empirical Bond Order (REBO) potential in classical Molecular Dynamics (CMD) and reverse non-equilibrium MD (RNEMD) simulations to investigate the mechanical and thermal transport properties of several GY monolayers \cite{wang2016}. They reported 22.9, 15.8, 32.8, 19.8, and 20.6 W/mK thermal conductivities for $\alpha$, $\beta$, $\gamma$, $\delta$, and 6,6,12-GY, respectively. Mortazavi and Zhuang also explored the thermal properties of some GY phases \cite{mortazavi2022}, using a machine learning interatomic potential (MLIP) derived from ab initio MD (AIMD) simulations. Their study reported 30.0, 69.0, 29.0, and 32.0 W/mK thermal conductivity for $\beta$, $\gamma$, $\delta$, and $\lambda$-GY, respectively.

The thermal transport of $\gamma$-GY has been further investigated by Zhan \textit{et al.}, who found a thermal conductivity of 31.4 W/mK \cite{zhan2014}, while Jing \textit{et al.} using a non-equilibrium MD (NEMD) simulations with a modified REBO potential incorporating longer-range Lennard-Jones interactions, reported a value of 64.8 W/mK \cite{jing2015}. Jiang and colleagues, using first-principles calculations, obtained a thermal conductivity of 76.4 W/mK for $\gamma$-GY \cite{jiang2017}. Other members of the GY family, such as graphdiyne and HGY, have also been investigated, with thermal conductivities ranging from 13.2 up to 29.3 W/mK \cite{jing2015,sajjad2023,mortazavi2023}.

These studies highlight the significant attention GY materials are receiving in the literature on carbon-based materials. Collectively, they reveal that the thermal conductivities of different GY monolayers are approximately two orders of magnitude lower than those of graphene \cite{mu2014,barbarino2015}. This decrease is primarily attributed to the presence of acetylene groups in GY structures \cite{soni2014,gao2019}. The reported thermal properties of GY materials, some already synthesized, illustrate their relevance in emerging thermal transport technologies.

A novel GY material, 8-16-4-GY, was introduced by Bandyopadhyay \textit{et al.} \cite{bandyopadhyay20218}. This initial study explored its dynamical, mechanical, thermal, and thermodynamic stability. Also, it provided an exact analytical expression for the generic dispersion relation of the square nodal-line semimetal class. The authors discussed the edge states associated with the quantized Zak phase, highlighting potential applications in acoustic crystals and other photonics-related technologies \cite{bandyopadhyay20218}.

Building on this, our group further characterized the 8-16-4-GY properties, which we renamed Sun-GY for simplicity. We demonstrated that Sun-GY's electronic band structure exhibits Dirac cones and remains structurally stable up to 10\% externally applied strain \cite{tromer2023mechanical}. In terms of optical properties, our findings revealed significant optical activity in the infrared, visible red, and ultraviolet regions. Additionally, we used Classical Molecular Dynamics (CMD) simulations to investigate its mechanical properties and heat capacity \cite{tromer2023mechanical}. Peng and collaborators extended this work by carrying out CMD simulations on larger-scale Sun-GY systems with defects and nanocracks, presenting a detailed analysis of its mechanical behavior \cite{peng2024effect}.

From another perspective, Jafari \textit{et al.} studied the influence of adsorbed 3d transition metals on Sun-GY, demonstrating that the material retained its metallic nature across all tested adsorbents. Moreover, the study showed that Sun-GY became magnetic when decorated with Cr, Mn, and Fe \cite{jafari2024evaluation}. Overall, Sun-GY has shown potential for a wide range of applications, drawing significant interest from the scientific community. However, its thermal transport properties remain unexplored and deserve further investigation.

In this work, we have investigated the thermal conductivity of Sun-GY, carrying out NEMD simulations with the second-generation REBO potential, which has been widely used to describe atomic interactions in 2D carbon systems. We explored how the phononic properties of this material affect its thermal conduction capacity, which was obtained as a function of system size, allowing the identification of ballistic and diffusive thermal transport regimes. The spectral analysis of the phononic properties of Sun-GY, including phonon dispersion and vibrational density of states (VDOS), was conducted to understand the underlying mechanisms of its thermal conductivity. This study provides a deeper understanding of the thermal transport of the GY family and contributes to better addressing the development of carbon-based nanomaterial applications.

\section{Methodology}

In this study, we employed the Large-scale Atomic/Molecular Massively Parallel Simulator (LAMMPS) package to carry out Molecular Dynamics (MD) simulations \cite{Thompson2022}, using the second-generation REBO potential to describe the interatomic forces \cite{Brenner2002}. This potential has been widely used to study phonon thermal transport of various 2D carbon systems \cite{ong2011,mortazavi2014,khan2015,wei2022}, including graphyne structures such as sheets \cite{wang2015,zhang2017}, nanoribbons \cite{ouyang2012,zhan2014}, nanotubes \cite{zhao2015,chen2017}, and heterojunctions \cite{zhan2014,wang2016}.

We also tested the optimized Tersoff potential \cite{Lindsay2010}, which performs well for graphene. However, we found it unsuitable for Sun-GY due to its inability to conserve energy. The REBO potential, in contrast, explicitly accounts for both $\sigma$ and $\pi$ interactions between carbon atoms, offering an improved description of carbon nanostructures that contain a mix of single, aromatic, and triple bonds, compared to the Tersoff potential \cite{Lindsay2010,wang2015}.

To test the accuracy of the REBO potential in capturing the atomic bonding structure of Sun-GY, we calculated its phonon dispersions by diagonalizing the dynamical matrix using the GULP (General Utility Lattice Program) software \cite{gale1997,gale2003}.

In all MD simulations, we integrated the equations of motion with a 0.5 fs timestep. The systems were initially thermalized using a Nosé-Hoover thermostat at 300 K for 100 ps. Periodic boundary conditions were applied along the $x$ and $y$ axes, with a vacuum buffer zone (larger enough to avoid spurious interactions) along the $z$ direction. Each system was relaxed at a finite temperature to reach zero stress along the periodic directions, with the stress along the $z$ direction automatically being zero. After equilibration, the thermostat was switched off, and the system evolved under microcanonical (NVE) conditions.

The thermal conductivity of Sun-GY was assessed using RNEMD simulations, which is based on the Muller-Plather approach \cite{Muller1997}. The concept of RNEMD is to inject a heat flux into the structure and determine the resultant temperature gradient. 

\begin{figure}[t!]
\begin{center}
\includegraphics[width=\linewidth]{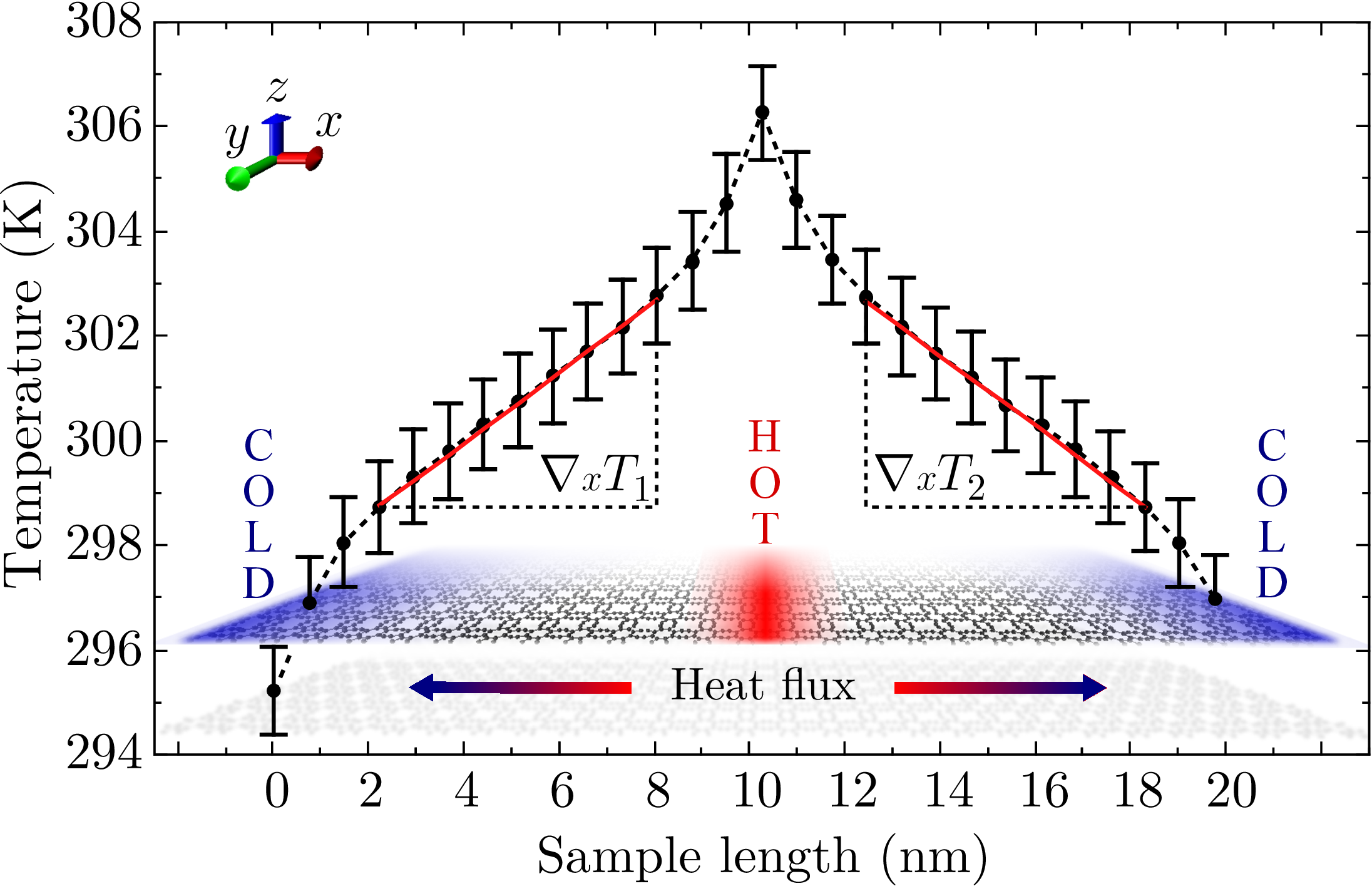}
\caption{Set up for the RNEMD method and typical temperature profile for a Sun-GY sheet with a nominal size of $20\ \text{nm} \times 20\ \text{nm}$ in the steady state. The heat flux is created along $x$-direction by exchanging the kinetic energy of slow-moving particles in the hot (red) region with fast-moving particles in the cold (blue) region. The data points with their respective error bars indicate the average temperature of each slab into which the system is divided, according to Eq. \eqref{eq:temperature}.}
\label{fig:rnemd-method}
\end{center}
\end{figure}

To create a heat flux, the system is divided into $n$ slabs along its length, and two of those slabs are chosen as ``hot'' and ``cold'' regions. In our setup, we chose the first slab as the cold region and the middle slab as the hot region. Due to periodic boundary conditions, slabs 0 and $N$ are the same.  Each slab had an average of 350 atoms. The heat flux is imposed by exchanging the kinetic energy of slow-moving particles in the hot region with fast-moving particles in the cold region, generating a temperature gradient, as shown in Figure \ref{fig:rnemd-method}. The kinetic energy swaps were performed every 500 simulation steps during 40 ns ($\sim 80 \times 10^6$ steps). The heat flux $J$, which is the energy transferred in a specific time $\Delta t$ through a surface perpendicular to the heat flux direction, is then given by \cite{Muller1997}:

\begin{equation}
    J(t)=\frac{1}{2 A \Delta t} \sum_\text{swaps} \frac{m}{2} \left( v_\text{hot}^2 - v_\text{cold}^2  \right) ,
\end{equation}

Where $m$ is the mass of the carbon atoms, $v_\text{hot}$ and $v_\text{cold}$ are the velocities of the faster-moving atoms in the cold slab and the slower-moving atoms in the hot slab, respectively. $A$ is the cross-sectional area of the sheet, which we defined as its width multiplied by its thickness. All simulated systems have the same nominal width of 20 nm along the $y$-direction, and we assume the same thickness of 0.335 nm along the $z$-direction \cite{ni2007graphene}. The factor 2 in the denominator is used because of the system's periodicity. 

At every simulation step the instantaneous temperature $T_i$ of $i$-th slab can be calculated from the equipartition theorem as \cite{Muller1997}

\begin{equation}
    \label{eq:temperature}
    T_i=\frac{2}{3 N_i k_B} \sum_{j \in i}^{N_i} \frac{m_j v_j^2}{2} ,
\end{equation}

where $N_i$ is the number of atoms in $i$-th slab, $k_B$ is Boltzmann’s constant, $m_j$ and $v_j$ are atomic mass and velocity of atom $j$, respectively. Thus, the temperature gradient is directly calculated from the average temperature in each region of the system. Figure \ref{fig:rnemd-method} shows the temperature profile for a sample with a nominal length of 20 nm. Clearly, the temperature profile shows good symmetry concerning the middle of the sample. The temperature profile is nonlinear near the hot and cold regions due to the finite size effects \cite{schelling2002}. Once the heat flux and the temperature gradient are stationary, we obtain the thermal conductivity for a sample of size $L$ directly from Fourier law \cite{Muller1997,pereira2022}:

\begin{equation}
    \label{eq:kappa-rnemd}
    \kappa(L)=-\frac{\langle J \rangle}{\left \langle \nabla_x T \right\rangle},
\end{equation}

where $\left \langle \nabla_x T \right\rangle$ is the arithmetic mean of the temperature gradient considering both directions of heat flux, as shown in
Fig. \ref{fig:rnemd-method}.

To analyze the vibrational spectrum of Sun-GY, we calculated its VDOS at room temperature. The VDOS was obtained by post-processing 100 ps trajectories in which atomic velocities were recorded every five fs. The VDOS was computed by calculating the Fourier transform of the velocity autocorrelation function, such that \cite{dickey1969,Felix2018}:

\begin{equation}
    \text{VDOS}(\omega)=\int_0^\infty \frac{\langle \textbf{v}(t)\cdot \textbf{v}(0) \rangle}{\langle \textbf{v}(0)\cdot \textbf{v}(0) \rangle} \exp{(-i\omega t)} dt,
\end{equation}

where $\textbf{v}$ is the atomic velocity, $\omega$ is the angular frequency, and $\langle \textbf{v}(t)\cdot \textbf{v}(0) \rangle$ is the velocity autocorrelation function (VACF), which is normalized such that $\text{VACF}(t=0)=1$.

\section{Results and Discussions}

\subsection{Phonon dispersion modeled by the second-generation REBO potential}

\begin{figure}[t!]
\begin{center}
\includegraphics[width=\linewidth]{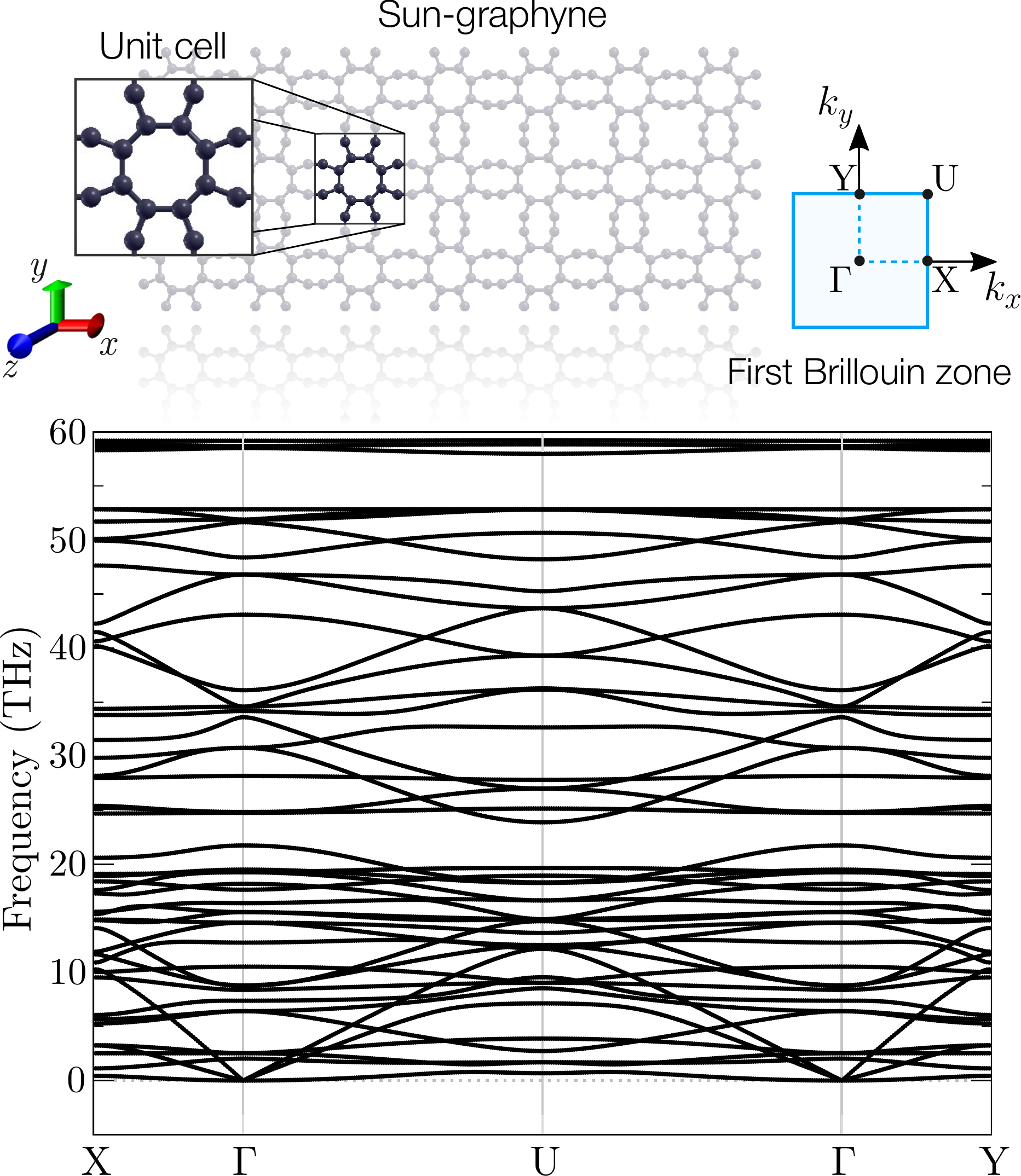}
\caption{Phonon dispersion for Sun-GY, modeled using the second-generation REBO potential. The absence of negative (imaginary) frequencies displays the stability of the structure with the chosen potential parameters.}
\label{fig:irida-disper}
\end{center}
\end{figure}

To ensure the accuracy of the second-generation REBO potential \cite{Brenner2002} in describing the atomic bonding structure of Sun-GY, we computed its phonon dispersion by diagonalizing the dynamical matrix using the lattice dynamics software GULP \cite{gale1997,gale2003}. Figure \ref{fig:irida-disper} shows the phonon dispersion along the high symmetry points of the Brillouin zone, derived from a unit cell containing 16 carbon atoms. The absence of phonon modes with negative (imaginary) frequencies confirms that Sun-GY's crystal structure is stable under the chosen potential.

The second-generation REBO potential accurately captures acoustic phonons but is less reliable for optical phonons \cite{ouyang2012,zou2016,wen2020}. Therefore, this potential is suitable for simulating thermal transport in our study, as acoustic phonons are the primary heat carriers in 2D materials \cite{lindsay2010_flexural,mu2014,barbarino2015,xu2015,taheri2021}. However, the thermal conductivity values obtained using this potential are expected to be lower than those predicted by machine-learning interatomic potentials or \textit{ab initio} methods that apply the Boltzmann transport equation \cite{arabha2021}. In Section \ref{Sec:spectrum}, we explore the implications of the phonon dispersion on the thermal transport properties of Sun-GY in more detail.

\subsection{Thermal conductivity's length dependence}

\begin{figure}[t!]
\begin{center}
\includegraphics[width=\linewidth]{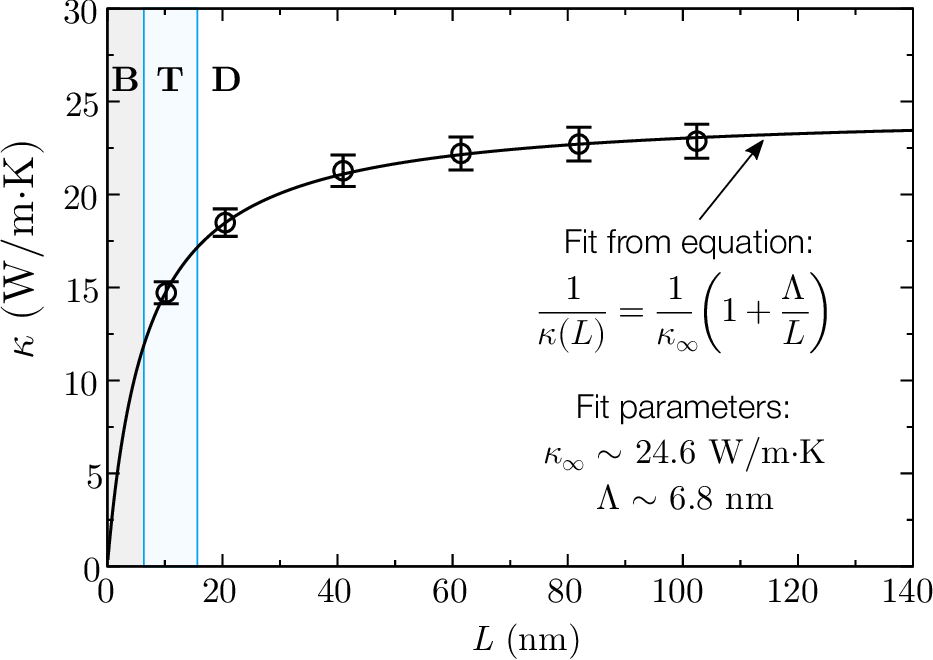}
\caption{Thermal conductivity of Sun-GY as a function of sample length. Data points from RNEMD simulations and lines from Eq. (\ref{eq:kappa-length}).
}
\label{fig:kappa}
\end{center}
\end{figure}

Thermal conductivity's length dependence is crucial for understanding fundamental phonon physics in low-dimensional systems and manipulating thermal transport through phonon engineering. Figure \ref{fig:kappa} illustrates the lattice thermal conductivity of Sun-GY as a function of system length, which can be explained by the kinetic theory of phonon transport \cite{schelling2002,chen2001}. The uncertainty in the data points was calculated using uncertainty propagation based on the uncertainties in average heat flux and temperature gradient, with an observed uncertainty of less than 5\% 

The behavior of $\kappa(L)$ reveals three distinct heat transport regimes, as shown in Figure \ref{fig:kappa}. The first one is the ballistic regime (region \textbf{B}, shaded gray), where $\kappa(L) \propto L$ for small system lengths, up to approximately 7 nm. In this region, the phonon mean free path exceeds the system length, making the thermal conductivity directly proportional to the size of the system.

For system lengths larger than 16 nm, we enter the diffusive regime (region \textbf{D}), where thermal conductivity shows only a weak dependence on system length. The phonon mean free path in this regime is shorter than the system length. Between these two regimes lies the ballistic-diffusive transition regime (region \textbf{T}, shaded blue), where the system length is comparable to the phonon mean free path. In this transition region, the dependence of thermal conductivity on system length decreases. In summary, phonons with intrinsic mean accessible paths longer than the system length travel ballistically. In contrast, those with shorter mean accessible paths are scattered diffusively.

Due to significant size effects arising from the limitation of the phonon mean free path to the region between the heat reservoirs, the conductivity for a system of length $L$ is expected to behave as \cite{schelling2002,felix2022}:

\begin{equation}
\label{eq:kappa-length}
    \frac{1}{\kappa(L)}=\frac{1}{\kappa_\infty}\left(1+\frac{\Lambda}{L}\right),
\end{equation}

where $\kappa_\infty$ represents the material's intrinsic (length-independent) conductivity. At the same time, $\Lambda$ denotes the effective mean free path of the heat carriers. By fitting this model to the simulation data from systems of increasing length, we can estimate both $\kappa_\infty$ and $\Lambda$ for Sun-GY. The estimated intrinsic lattice thermal conductivity of Sun-GY at room temperature is approximately 24.6 W/mK. At the same time, the effective phonon mean free path is about 6.8 nm. Using the method recently proposed in reference \cite{tromer2023}, we estimated the lattice thermal conductivity of sun-GY to be approximately 30 W/mK. This value agrees with the results obtained via the RNEMD method discussed in this work.

It is important to note that this extrapolation method is accurate only if the smallest system size used in the RNEMD simulations is comparable to the longest mean accessible paths of the phonons that dominate thermal transport \cite{sellan2010,Felix2020}. The fitting performed using Eq. (\ref{eq:kappa-length}) aligns very well with the data points from the RNEMD simulations, demonstrating its predictive solid capability.

The intrinsic thermal conductivity we obtained for Sun-GY is of the same magnitude as other structures in the graphyne family, as shown in Table \ref{tab:conductivity}. However, it is at least two orders of magnitude lower than the approximately 1200 W/mK reported for pure graphene using the same empirical potential \cite{mu2014,barbarino2015,felix2024irida}. The presence of acetylenic bonds is known to impact the lattice thermal conductivity of other graphyne systems significantly \cite{ouyang2012,zhang2012,zhan2014,soni2014,tan2015,wang2015,zhang2017,gao2019}. We attribute the lower $\kappa_\infty$ value of Sun-GY compared to graphene to its acetylenic bonds, which increase the scattering of heat carriers and thus reduce its thermal conductivity.

\begin{table*}[t!]
\caption{Calculated intrinsic lattice thermal conductivity at room temperature for some structures of the graphyne family.}
\begin{center}
\begin{tabular}{c c c c}
\hline \hline
Graphyne	&	$\kappa_\infty$ (W/mK)	&	Method	&	Reference \\
\hline\hline
Sun	                        &	24.6	&	RNEMD -- REBO potential	&	This work	\\\hline
$\alpha$	                &	22.9	&	RNEMD -- REBO potential	&	Wang \textit{et al.} (2016) \cite{wang2016}	\\\hline
\multirow{2}{*}{$\beta$}	&	15.8	&	RNEMD -- REBO potential	&	Wang \textit{et al.} (2016) \cite{wang2016}	\\
                    	  &	  30.0	  &	  NEMD -- ML potential	  &	Mortazavi \& Zhuang (2022) \cite{mortazavi2022}	\\\hline
\multirow{5}{*}{$\gamma$}	&	32.8	&	RNEMD -- REBO potential	&	Wang \textit{et al.} (2016) \cite{wang2016}	\\
                            &	69.0	&	NEMD -- ML potential	&	Mortazavi \& Zhuang (2022) \cite{mortazavi2022}	\\
                            &	31.4	&	RNEMD -- REBO potential	&	Zhan \textit{et al.} (2014)\cite{zhan2014}	\\
                            &	64.8	&	NEMD -- AIREBO potential	&	Jing \textit{et al.} (2015) \cite{jing2015}	\\
                            &	76.4	&	\textit{ab initio} calculation 	&	Jiang \textit{et al.} (2017) \cite{jiang2017}	\\\hline
\multirow{3}{*}{$\delta$}	&	19.8	&	RNEMD -- REBO potential	&	Wang et al. (2016) \cite{wang2016}	\\
                            &	19.4	&	RNEMD -- REBO potential	&	Zhang \textit{et al.} (2017) \cite{zhang2017}	\\
                            &	29.0	&	NEMD -- ML potential	&	Mortazavi \& Zhuang (2022) \cite{mortazavi2022}	\\\hline
$\lambda$                   &	32.0	&	NEMD -- ML potential	&	Mortazavi \& Zhuang (2022) \cite{mortazavi2022}	\\\hline
6,6,12                      &	20.6	&	RNEMD -- REBO potential	&	Wang \textit{et al.} (2016) \cite{wang2016}	\\\hline
\multirow{2}{*}{Holey}      &	29.3	&	\textit{ab initio} calculation 	&	Sajjad \textit{et al.} (2023) \cite{sajjad2023}	\\
                            &	14.0	&	NEMD -- ML potential	&	Mortazavi (2023) \cite{mortazavi2023}	\\\hline
Graphdiyne                   &	13.2	&	NEMD -- AIREBO potential	& Jing \textit{et al.} (2015)	\cite{jing2015}	\\
\hline
\hline
    \end{tabular}
    \label{tab:conductivity}
\end{center}
\end{table*}

\subsection{Spectral analyses}
\label{Sec:spectrum}

To gain a deeper understanding of the underlying mechanism responsible for the decrease in the thermal conductivity of Sun-GY compared to pristine graphene, Figure \ref{fig:spectrum} provides for both (a) Sun-GY and (b) graphene: (I) a spectral analysis of the dispersion relation along the heat flow direction, (II) the phonon group velocity, and (III) the vibrational density of states at room temperature.

\subsubsection{Phonon dispersion}

\begin{figure*}[ht!]
\begin{center}
\includegraphics[width=0.7\linewidth]{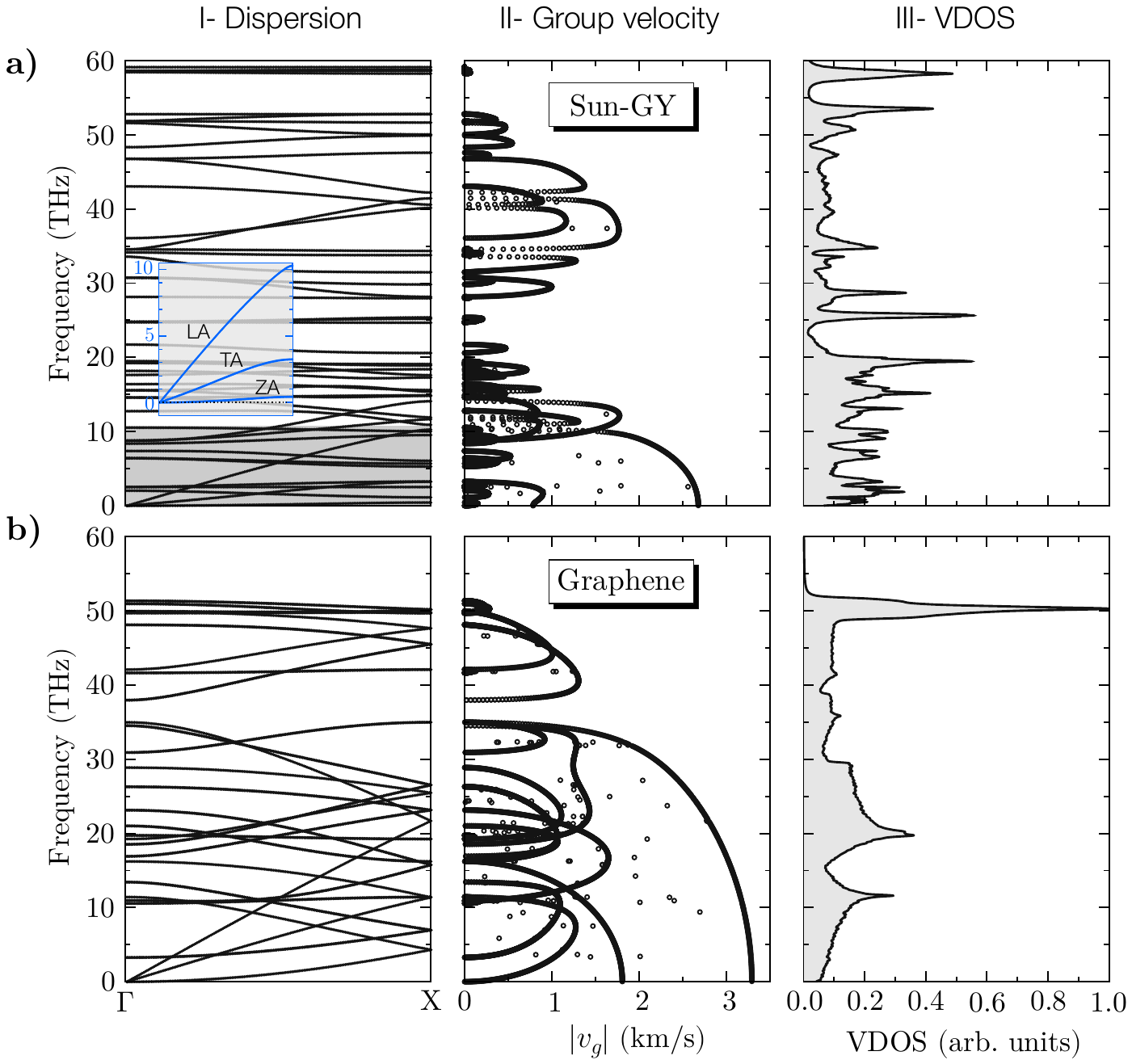}
\caption{(I) Spectral analysis of the dispersion relation along the $\Gamma\rightarrow$ X path, (II) phonon group velocity, and (III) vibrational density of states at room temperature for both (a) Sun-GY and (b) graphene.}
\label{fig:spectrum}
\end{center}
\end{figure*}

To systematically compare the phononic properties of Sun-GY and graphene, we calculated the dispersion relations for both materials using unit cells with 16 atoms and similar dimensions. We focused on the $\Gamma \rightarrow$ X path, as it aligns with the direction of thermal transport. Figures \ref{fig:spectrum} (a -- I) and (b -- II) display the dispersion relations for Sun-GY and graphene, respectively.

The in-plane longitudinal (LA) and transverse (TA) modes show linear dispersion among the three acoustic phonon modes in both materials. However, the out-of-plane flexural (ZA) mode exhibits quadratic dispersion around the $\Gamma$ point, a characteristic feature of 2D materials. Notably, the dispersion branches for Sun-GY are less steep than those of graphene. The inset plot in Figure \ref{fig:spectrum} (a -- I), which highlights only the acoustic modes of Sun-GY, shows that the LA, TA, and ZA branches reach frequencies of approximately 10, 3, and 0.5 THz, respectively. In contrast, for graphene (considering band folding), the LA, TA, and ZA branches reach frequencies of around 35, 16, and 11.5 THz at the X point, respectively.

Additionally, the optical modes in Sun-GY are generally flatter than those in graphene. The first optical phonon mode of Sun-GY has a shallow frequency of approximately 2 THz, which can lead to increased phonon-phonon scattering due to interactions between acoustic and optical modes \cite{lindsay2010_flexural}. Specifically, the quadratic dispersion of the ZA mode in Sun-GY results in a high density of available states at low energies, allowing significant interactions with low-frequency optical modes.

\subsubsection{Group velocity}

The group velocity reveals one of the main implications of phonon dispersion for thermal transport. The phonon group velocity is closely related to thermal conductivity because it determines how fast heat is transferred along the system \cite{ziman2001}. In general, the phonon group velocities are given by:

\begin{equation}
\label{eq:gvel}
    v_g (\omega) = \frac{d\omega}{dq}
\end{equation}

where $\omega$ represents the frequency of a given mode, and $q$ denotes the wave vector. Given that the dispersion curves of Sun-GY are less steep compared to those of graphene, it follows from Eq. (\ref{eq:gvel}) that Sun-GY is expected to have a lower group velocity.

Figures \ref{fig:spectrum} (a -- II) and (b -- II) show the absolute value of the phonon group velocities as a function of frequency for Sun-GY and graphene, respectively. The data indicate that the phonon group velocities for Sun-GY are smaller than those for graphene across most frequencies in the phonon spectrum. The average group velocities are approximately 356 m/s for Sun-GY and 635 m/s for graphene. Since the lattice thermal conductivity is proportional to the square of the group velocities \cite{ziman2001}, this difference highlights the lower thermal conductivity of Sun-GY compared to graphene. These findings are consistent with previous studies on the thermal transport properties of other graphyne structures \cite{soni2014,tan2015,gao2019}.

For instance, it is well-known that the phonon dispersion relations in the graphyne family are flatter and thus exhibit lower group velocities \cite{wang2015}. This trend is due to the weak bonding in the acetylenic linkages \cite{ouyang2012,zhang2012,zhan2014,soni2014,tan2015,wang2015,zhang2017,gao2019}. Specifically, Tan \textit{et al.} reported that the presence of $sp$ bonds reduces the group velocities of the acoustic branches and introduces new optical branches, leading to a significantly lower phonon thermal conductivity in the graphyne family compared to graphene \cite{tan2015}.

\subsubsection{Vibrational density of states}

Figure \ref{fig:spectrum} clearly shows that the vibration characteristics of the carbon atoms of Sun-GY [panel (a -- III)] differ from those of graphene [panel (b -- III)].
For instance, there are more prominent peaks in the VDOS of Sun-GY than in graphene.
Comparing the frequency positions of these peaks with the phonon dispersion relations [panel (a -- I)], we find that these VDOS peaks are van Hove singularities \cite{van1953}, where the group velocity of phonons tends to zero  [see panel (a -- II)]. 
In general, the phonon modes near these singular peaks are expected to be highly localized for the graphyne family due to the presence of acetylenic bonds \cite{wang2015}.

\begin{figure}[t!]
\begin{center}
\includegraphics[width=\linewidth]{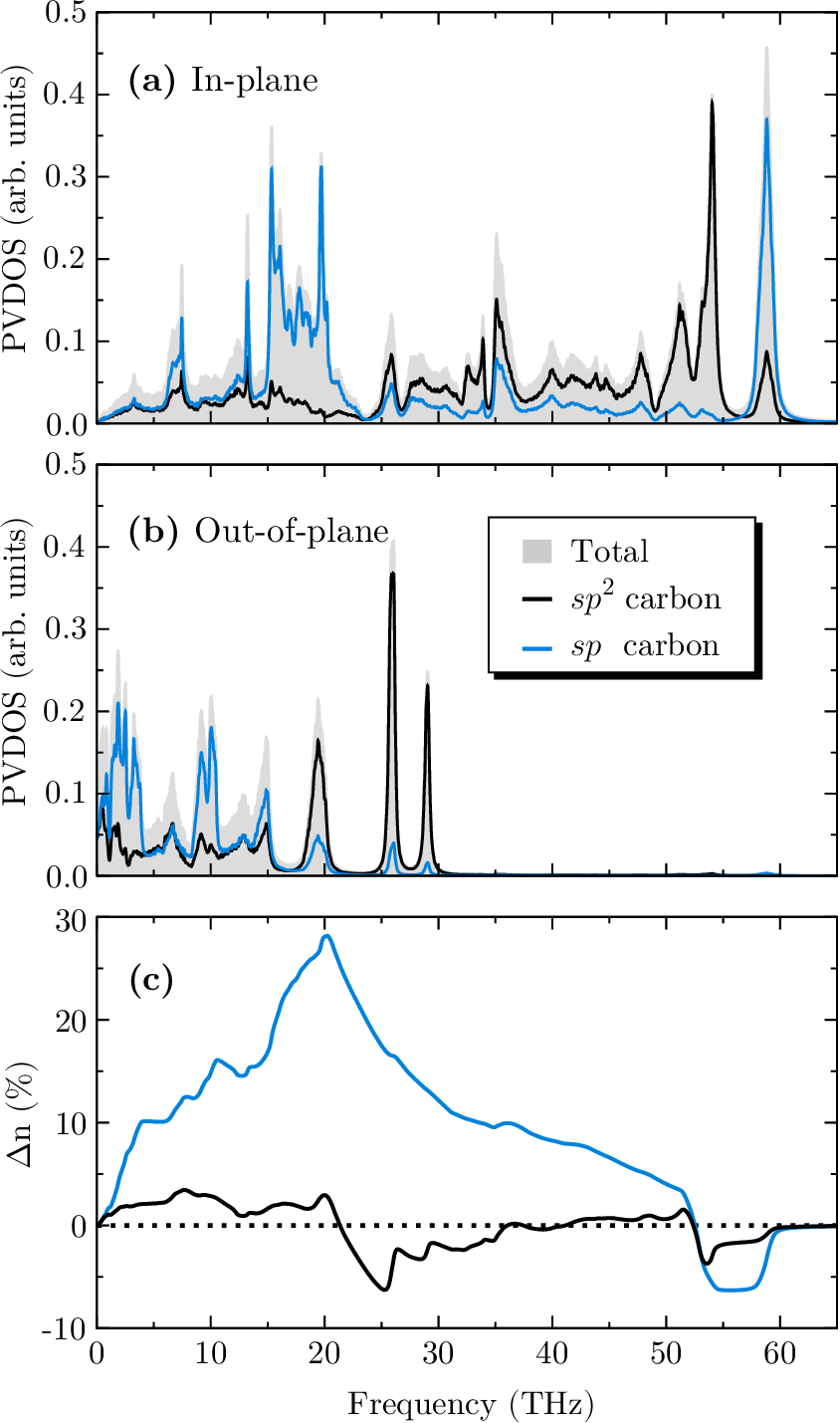}
\caption{Simulated projected vibrational density of states (PVDOS) for the (a) in-plane and (b) out-of-plane phonon modes of Sun-GY. (c) displays the changes in Sun-GY phonon populations relative to graphene. The gray baseline shows the total contribution of the VDOS; the black line shows the contribution of the carbon atoms with $sp^2$ hybridization that form the octagons, and the blue line shows the contribution of the carbon atoms with $sp$ hybridization that form the acetylenic linkages. 
}
\label{fig:pvdos}
\end{center}
\end{figure}

To carefully analyze the VDOS of Sun-GY, Figure \ref{fig:pvdos} presents the projected vibrational density of states (PVDOS) for the in-plane [panel (a)] and out-of-plane [panel (b)] phonon modes. The gray baseline shows the total contribution. The black line shows the contribution of the carbon atoms with $sp^2$ hybridization that form the octagons. The blue line shows the contribution of the carbon atoms with $sp$ hybridization that form the acetylenic linkages. 

Note that out-of-plane (ZA) phonon modes contribute the most to the VDOS in the low-frequency regime (below 11 THz), where acoustic phonons dominate and are likely the primary heat carriers in Sun-GY. Low-frequency phonons with longer wavelengths typically contribute more to heat transport in nanomaterials. These heat carriers undergo fewer scatterings and travel longer distances throughout the material. Therefore, they usually carry significant energy and can contribute more effectively to thermal transport. Moreover, it has been recognized that the flexural polarization (ZA) due to out-of-plane phonon modes is fundamental in the thermal transport of graphene \cite{lindsay2010_flexural,seol2010,pereira2013,barbarino2015} and other 2D materials \cite{taheri2021}.

Figure \ref{fig:pvdos} provides insights into the individual contributions of $sp$ and $sp^2$ hybridized carbon atoms to the vibrational density of states (VDOS) of Sun-GY. We first examine the contributions of $sp$ and $sp^2$ carbon atoms to the in-plane phonon modes. In this way, Figure \ref{fig:pvdos} (a) shows that $sp$ hybridized carbon atoms predominantly contribute to high-frequency isolated phonon modes around 60 THz. This observation is consistent with other theoretical predictions \cite{popov2013,soni2014,perkgoz2014,tan2015,gao2019,serafini2022}. Experimental studies \cite{li2010} indicate that $sp$ carbon bonds correspond to the stretching vibrations of triple bonds ($\mathrm{-C\equiv C-}$). Additionally, $sp$ carbon atoms significantly influence phonons with frequencies below 20 THz. In comparison, $sp^2$ carbon atoms mainly contribute to intermediate frequency phonons ranging from 20 to 55 THz.

By revisiting Figure \ref{fig:spectrum} (a--II), the lower group velocity observed in the Sun-GY sheet compared to graphene can be attributed to the contributions of $sp$ hybridized carbon atoms. In summary, the presence of acetylenic bonds is a critical factor in decreasing the thermal transport of phonons in Sun-GY, a trend also observed in other members of the graphyne family \cite{ouyang2012,zhang2012,zhan2014,soni2014,tan2015,wang2015,zhang2017,gao2019}. This feature underscores the importance of understanding the vibrational properties of carbon-based materials in materials science.

Figure \ref{fig:pvdos} (c) shows the changes in phonon populations of Sun-GY relative to graphene, which were calculated from the ratio between the occupation of phonon modes, defined as \cite{Felix2018}:

\begin{equation}
    \Delta\text{n} = \displaystyle\frac{1+\displaystyle\int_0^\omega \text{VDOS}_\text{Sun-GY} \ d\omega'}{1+\displaystyle\int_0^\omega \text{VDOS}_\text{Graphene} \ d\omega'}\ .
\end{equation}

Here, $\Delta \text{n} (\omega)=0$ corresponds to no change in phonon population relative to graphene, $\Delta \text{n} (\omega)<0$ indicates a decrease, and $\Delta \text{n} (\omega)>0$ indicates an increase in phonon population. Note that the phonon population changes in Sun-GY for $sp^2$ carbon atoms are always less than 6\%. On the other hand, $sp$ carbons induce a progressive increase in the phonon population up to 20 THz, which can reach 30\% compared to graphene. Since this increase in the phonon population coincides with the decrease in the group velocity, we believe it might be associated with phonon localization due to acetylenic bonds. Similar to isotopes or vacancies \cite{zhang2010,xie2023}, these localized phonon modes acting as impurities can significantly increase the phonon scattering rate, shortening the phonon's lifetime and decreasing thermal transport \cite{wang2015}.

\section{Summary and Conclusions}

The thermal transport properties of Sun-GY, a 2D carbon allotrope, were explored through molecular dynamics simulations. We employed the second-generation REBO potential to model the interactions between carbon atoms, which effectively captured the bonding characteristics of Sun-GY. We utilized reverse non-equilibrium molecular dynamics (RNEMD) simulations to determine the thermal conductivity, providing detailed insights into heat transport and its dependence on system length. Our results reveal three distinct thermal transport regimes: a ballistic regime at shorter lengths; a diffusive regime at longer lengths; a ballistic-diffusive transition regime at intermediate lengths.

The intrinsic thermal conductivity of Sun-GY was estimated to be approximately 24.6 W/mK. This value is significantly lower than graphene's but aligns with the thermal conductivities found for other graphyne systems. The presence of acetylenic linkages in Sun-GY increases the scattering of heat carriers, contributing to the lower thermal conductivity.

Spectral analysis of phonon dispersion and vibrational density of states revealed that Sun-GY has flatter optical modes and lower phonon group velocities than graphene. These findings are consistent with the observed lower thermal conductivity in Sun-GY. The acetylenic linkages disrupt heat transport by reducing phonon group velocities and enhancing the coupling between acoustic and optical modes.

Overall, this study confirms that the REBO potential is suitable for simulating thermal transport in graphyne materials and provides a deeper understanding of Sun-GY's thermal properties. The reduced thermal conductivity of Sun-GY, as revealed by our research, opens up exciting possibilities for its application in various fields. From thermal management systems for electronic devices to thermal barrier coatings in aerospace engines and insulation materials in construction, Sun-GY's unique properties can significantly enhance performance and efficiency, inspiring further research and innovation.

\begin{acknowledgments}
This work received partial support from Brazilian agencies CAPES, CNPq, and FAPDF.
I.M.F. thanks the National Council for Scientific and Technological Development (CNPq) for grant no. 153604/2024-7.
R.M.T. acknowledges the support of the MackGraphe.
L.D.M. acknowledges the support of the High Performance Computing Center at UFRN (NPAD/UFRN).
D. S. G. acknowledges the Center for Computing in Engineering and Sciences at Unicamp for financial support through the FAPESP/CEPID Grant \#2013/08293-7.
L.A.R.J. acknowledges the financial support from FAP-DF 00193.00001808/2022-71 and 00193-00001857/2023-95 grant and FAPDF-PRONEM grant 00193.00001247/2021-20, and CNPq grant 350176/2022-1.
M.L.P.J. acknowledges the financial support of the FAP-DF grant 00193-00001807/2023-16. Authors also acknowledge support from CENAPAD-SP (National High-Performance Center in São Paulo, State University of Campinas -- UNICAMP, projects proj950 and proj960) and NACAD (High-Performance Computing Center, Lobo Carneiro Supercomputer, Federal University of Rio de Janeiro -- UFRJ, project: a22002) for the computational support provided. 
\end{acknowledgments}

\appendix

\nocite{*}

\bibliography{manuscript.bib}
\end{document}